# Elastic properties of polycrystalline $YBa_2Cu_3O_{7-\delta}$: Evidence for granularity induced martensitic behavior

**C. Stari[*] and A. Moreno-Gobbi**

*Instituto de Física, Facultad de Ciencias, Universidad de la República,*
*Iguá 4225 - C.P. 11400, Montevideo, Uruguay*

**A. W. Mombru**

*Instituto de Física, Facultad de Química, Universidad de la República,*
*Isidoro de María 1620, C.P.11800, Montevideo, Uruguay*

**S. Sergeenkov**

*Bogoliubov Laboratory of Theoretical Physics, Joint Institute for Nuclear Research,*
*Dubna, 141980 Moscow Region, Russia*

**A. J. C. Lanfredi, C. A. Cardoso and F. M. Araujo-Moreira[**]**

*Department of Physics and Physical Engineering, Materials and Devices Laboratory,*
*Multidisciplinary Center for Development of Ceramic Materials-UFSCar, Caixa Postal 676,*
*São Carlos SP, 13565-905 Brazil*

**Abstract**

In this work we present the study of the elastic properties of polycrystalline samples of superconducting $YBa_2Cu_3O_{7-\delta}$ prepared by the *sol-gel* method. The quality of all samples was checked by x-ray diffraction and scanning electronic microscopy while their physical properties were verified by transport and magnetic measurements. The elastic study was performed using the standard *pulse-echo* technique through measuring the phase velocity and the attenuation of ultrasonic waves (in the range of a few MHz) as a function of temperature. We have focused this study on the low temperatures interval (T < 200K). The obtained results show a strong hysteretic behavior in the ultrasonic attenuation (in addition to usually observed hysteretic behavior for the velocity) which strongly supports the existence of a *martensitic-like* phase above the superconducting critical temperature $T_C$. We argue that this peculiar behavior can be attributed to the granularity present in the samples.

---

[*] Present Address: Instituto de Física, Facultad de Ingenieria, Universidad de la República, J. Herrera y Reissig 565, Casilla de Correo N. 30, Código Postal: 11300 Montevideo, Uruguay

[**] Corresponding author; e-mail address: faraujo@df.ufscar.br

# 1. Introduction

As is well known, in normal metals, an important contribution to the ultrasonic attenuation arises from the interaction of the ultrasonic waves with the conduction electrons. When the normal metal turns superconducting (for temperatures below some critical value, $T_C$), this contribution suddenly decreases and goes to zero for T = 0 [1]. Over the last years, the ultrasound technique has been used for better understanding of superconductivity mechanisms through revealing details of gap and band anisotropies in the high-$T_c$ (HTS) copper oxides in which the superconductivity is believed to reside within the two-dimensional $CuO_2$ planes and the energy gap is known to exhibit $d_{x^2-y^2}$ symmetry[2]. Recently, interesting results regarding structure sensitive effects of oxygen removal on ultrasonic attenuation and elastic behavior of HTS has been reported[3]. Also, ultrasonic experiments performed under pressure in superconducting $MgB_2$, have provided valuable information about its elastic properties as well as thermodynamic parameters. This type of experiments is important to understand the mechanisms of HTS and the evolution of the critical temperature with applied pressure, giving valuable clues for guiding chemical substitutions and finding new superconducting materials[4]. At the same time, Lupien and co-workers[5] have used the intrinsic angular resolution of ultrasound attenuation to directly probe the anisotropy of the quasi-particle excitation spectrum of superconducting $Sr_2RuO_4$ in zero magnetic field. As was emphasized by Moreno and Coleman[6], in anisotropic superconductors transverse ultrasound attenuation provides a weakly coupled probe of momentum current correlations in electronic systems. More recently, Kumar *et al.*[7] have performed resonant ultrasound measurements to characterize the anisotropic elastic properties of $CeRhIn_5$. Also, Joynt and co-workers[8] have studied the superconducting phases of the heavy-fermion compound $UPt_3$, which is the first compelling example of a superconductor with an order parameter of unconventional symmetry. Those authors have determined in details the phase diagram of $UPt_3$ in the H-T plane by ultrasound and thermodynamic measurements while their experiments under pressure provided clear evidence for a strong coupling between antiferromagnetism and superconductivity in this compound.

Martensitic transformations are considered as diffusionless solid-solid phase transitions. They have been detected over the years in many types of materials and using several different techniques[9,14]. In particular, a specific group of metals, the so-called *shape memory alloys* (found to exhibit pronounced martensitic transformations) has been intensively studied basically by using mechanical spectroscopy. At the same time, martensitic transformations in non-metallic materials (like ceramic superconductors) are studied by using a more sensitive ultrasound technique (one can find just few references in the literature[9,10,15] on the subject matter).

In the present work, we use the powerful ultrasound technique to study the elastic behavior of ceramic samples of superconducting $YBa_2Cu_3O_{7-\delta}$ (YBCO) prepared by the *sol-gel* method. We have also prepared some samples by the standard solid state reaction to establish the differences between some of their physical and structural features. The elastic study was performed using the standard *pulse-echo* technique by measuring the phase velocity and the attenuation of ultrasonic waves, as a function of temperature.

## 2. Samples Characterization and Experimental Procedure

As was pointed out before, polycrystalline samples of superconducting YBCO were prepared by using two different chemical routes: the standard solid state reaction method (labeled as samples M1) and the *sol-gel* method (labeled as samples M2)[16,17]. The structural quality of all samples was verified through both X-ray diffraction (XRD) and scanning electron microscopy energy dispersive spectroscopy (SEM-EDS). Those experiments reveal that sample M1 is of higher structural quality than sample M2 showing a more homogeneous structure with grains similar in size and shape (see Figure 1). On the other hand, sample M2 presents a broad distribution of grain parameters as well as higher porosity than sample M1. The phase purity and structural characteristics of our samples were confirmed by Rietveld analysis[18] of the x-ray powder diffraction data, obtained by using Cu-K$\alpha$ radiation with a Seifert Scintag PAD-II diffractometer (with 2$\theta$ range from 20º to 100º). The Rietveld analysis was performed employing the FULLPROF set of programs[19].

The physical properties of all samples were verified by transport and magnetic measurements. Samples were shaped as cylinders with diameter of about 10.0 mm and 8.0 mm height. The upper and bottom faces of the samples were carefully polished to reach the necessary parallelism between them (less than 1 µm) in order to guarantee the correct performance of the ultrasound technique. The study of the elastic properties of our samples was performed using the standard *pulse-echo* method[20] by measuring the phase velocity and the attenuation of ultrasonic waves (propagating inside the sample) as a function of the absolute temperature. This method allows detecting small variations of the phase velocity with high accuracy by determining the ultrasound wave traveling time in a round trip, $\Delta t$, obtained between two successive echoes. Namely, if $d$ is the height of the sample, the ultrasound velocity is given by $v=2d/\Delta t$. Due to different effects like internal reflections and diffraction, and also due to the presence of different media with different values of the ultrasound impedance, there will inevitably appear shifts in the phase of the ultrasonic wave. To overcome these problems and to accurately determine a specific point in the phase cycles, we have used the Papadakis method[21]. In ideal conditions, this method allows to determine the absolute traveling time of the ultrasound wave with an error smaller than 1 *ppm*. However, for most common experimental situations[22] its typical value can be around one part in $10^4$. The experimental arrangement consists mainly of a MATEC generator/receptor and a MATEC attenuation recorder of ultrasonic pulses connected to a digital TEKTRONIX oscilloscope and to a digital KEITHLEY multimeter. Low temperatures where reached by using a JANIS close-cycle cryogenic system and a LAKE SHORE temperature controller. Experiments were performed using 5 MHz quartz X and Y transducers. To couple sample and transducer we have used both *Nonac* (for longitudinal and transverse waves, at low temperature) and *Salol* (for transverse waves, at temperatures higher than 200K). The experiment was fully automatized and the temperature was swept at a rate of approximately 0.5 K/min (for both warming-up and cooling-down processes).

## 3. Results and Discussion

After performing the structural characterization of samples mentioned in the previous section, we performed the study of both electrical and magnetic properties. First of all, using the well-known four-contacts method (performed in a Janis cryostat coupled to a Helium close-cycle refrigerator) we have measured the resistance of the sample as a function of temperature. On the other hand, by using a Quantum Design MPMS magnetometer we have also measured the magnetic response of our samples by determining the complex AC susceptibility as a function of temperature (for sample M1) and magnetization as a function of temperature (for sample M2). These results are shown in Figs. 2 and 3 respectively.

Turning to the discussion of the obtained results, we notice that the resistivity versus temperature curve shows a rather broad transition indicative of extrinsic (grain boundaries) and/or intrinsic (inhomogeneities in the oxygen content) granularity. Despite the fact that the oxygen content of both samples is almost the same (0.15 for M1, and 0.14 for M2), its distribution in sample M2 is less homogeneous than in sample M1. As expected, this feature is also observed in the magnetic characterization of the samples (see Fig. 3). Transport and magnetic experiments performed on M1 sample show sharper superconducting transition, starting around T=92K (and with width of about 5K).

The elastic study was performed by determining the velocity and the attenuation of ultrasound waves of frequency of 5 MHz, and they have been analyzed in both modes, longitudinal and transverse. From the obtained results for sample M2 (see Fig. 5), a strong hysteretic behavior is seen between warming-up and cooling-down experiments (both performed at a rate of 0.5 K/min). At the same time, sample M1 does not exhibit any visible hysteretic behavior (see Fig. 4).

In what follows, we separately discuss the hysteretic behavior of velocity and attenuation in sample M2 for low and high temperatures.

All the obtained results for the high-temperature interval (T > 200 K) show no hysteretic behavior, neither in attenuation nor for velocity experiments as a function of temperature. These results are consistent with those reported in the literature [23-26].

On the other hand, for the low-temperature interval (T < 200 K), transport measurements point out that the superconducting transition temperature is around 90 K. Over this low-temperature region, we have found a very complex elastic behavior, where there are many superimposed mechanisms responsible for the elastic properties. Both attenuation and velocity curves have strong hysteretic behavior when warming-up and cooling-down the sample, in this interval. The superconducting transition temperature, $T_C$, is found to lie around the middle of this interval. The attenuation presents a broad and asymmetric peak centered closely to $T_C$. This peak appears in both warming-up and cooling-down experiments, and is shifted by approximately 15K to lower temperatures during the cooling process. This behavior, observed for longitudinal and transverse ultrasound waves, is remarkably different from that observed in the high-temperature interval. It is also seen that the low-temperature peaks are superimposed to an almost linear component in temperature, which is indicative of the existence of an extra mechanism responsible for the elastic properties in this region. When we carefully analyze Fig. 6, we can see a second peak with smaller amplitude, close to the largest and central one. Also, this central peak is asymmetric and very broad (broadness higher than 100K), and its asymmetry suggests the existence of a third peak, since just two peaks can not be responsible for such behavior. Curves with these characteristics, exhibiting broad peaks, are commonly associated with a discrete *multipeak* structure. This behavior is more evident when warming-up the sample. Even though our results do not show the existence of many peaks when cooling-down the sample, it is possible to observe that, after the maximum (close to 90K), there is a change in the concavity of the curve (around 66K), corresponding to a second peak, similar to that found when warming-up the sample. Even though this behavior of the attenuation is in general agreement with the observed by other authors[9,15,24] in the same material, there are still some differences as far as the quantity, localization and possible origins of those peaks are concerned. Typically, the results from the literature where the attenuation presents a peak structure are associated to measurements with oscillations in the low-frequency range (of the order of a few kHz[9,15,24]). For example, Cannelli et al.[15] and Cheng et al.[9] reported a similar peak structure for samples with similar oxygen content showing a double peak close to the transition temperature. On the other hand, Bhattacharya and co-workers[24] have reported attenuation measurements obtained from experiments performed at 5 MHz, where there is a peak around 65K, far away from $T_C$. In that work, the superconducting transition temperature appears in a plateau of the attenuation curve, which seems to be in contradiction with other data from the literature where the peak close to $T_C$ is commonly found.

It is important to notice that the strong hysteretic behavior of M2 sample observed in both the attenuation and the velocity is very similar to that presented in materials exhibiting a martensitic transformation[11-14]. Another peculiar behavior of these types of transitions is that the hysteretic feature occurs only when the sample is cooled below a certain temperature. In this regard, it is interesting to mention that Cannelli *et al.*[15] found that if the sample is cooled to a temperature higher than some critical value, the attenuation is not modified when the sample is warmed-up. However, if the sample is cooled down to a temperature below that critical value, then the attenuation will change to a hysteretic behavior when the sample is warmed up.

In our case the martensitic phase transformation starts at $T = M_s$ (see Figure 7) and ends at $T = M_F$, for the cooling-down process. On the other hand, when the sample is warmed-up, the transformation occurs between $T = A_S$ and $T = A_F$. We have chosen the beginning of the transformation at the temperature where the velocity starts to visibly change its behavior. As is seen in Figure 7, the beginning of the phase transformation coincides with the change in the derivative of the magnetic moment and resistivity curves. Therefore, the martensitic phase transition starts at $T = 65$ K for the cooling-down process, and at around $T = 110$ K when warming-up the sample.

Based on the above findings, we proceed to analyze the low-temperature peak in the attenuation at high frequencies. In doing so, we have decomposed it into several individual peaks. We have found that, assuming the existence of four peaks, it is possible to correctly describe all the experimental data. On the other hand, we have found that it is not possible to obtain such a result by considering just two or three peaks. To perform the decomposition of the experimental peak, we assume that the peaks in the attenuation have a Lorentz-like behavior of the form $\alpha(T) = (2\omega A/\pi)[4(T-T_0)^2 + \omega^2]^{-1}$ where $\omega$ is the width of the peak, A is the area under the curve, and $T_0$ is the peak temperature (Figure 8). The fitting parameters are described in Table 1. It is worth noting the preliminary subtraction of the linear contribution from the attenuation, clearly seen in Figure 7. Since it is the same for both cooling and warming processes, it should be associated with the same mechanism and hence can be safely subtracted.

Following this procedure, we obtain a peak (localized at a temperature below $T_C$) which appears in both cooling-down and warming-up processes. Also, a second peak appears at the beginning of the superconducting transition that is also present in both cooling and warming processes. A third peak appears at the beginning of the martensitic transformation, at around 60 K when cooling-down the sample, and at around 110K when warming it up. Finally, we obtain a fourth peak at higher temperatures also present in both thermal processes.

It is worthwhile to mention that there has been a long-standing discussion[9-10] on whether (or not) a martensitic transformation induces a superconducting transition in HTS. From our own results we can conclude that the beginning of the superconducting transition certainly occurs before the martensitic transformation which is most likely associated with extra stress due to grain sizes and irregular shapes.

At the same time, for sample M1, the velocity and attenuation results show neither hysteretic behavior nor peak structure associated with the martensitic phase transition. Previously, Mamsurova et al.[27] have attributed both the hysteretic behavior and the high-temperature peak in the attenuation to grain size and the density of the sample. Those authors have found that both features disappear when grain size is diminished together with an increase in the density of the sample. However, our second sample still exhibits a pronounced and well defined peak in the attenuation at high temperatures. These results could indicate that the attenuation peak at high temperatures as well as the hysteresis cycle are associated with different phenomena or, alternatively, that the peak in the attenuation is more sensitive to the microstructure of the sample. On the other hand, the hysteretic behavior observed in both the velocity and the attenuation (martensitic transformation) seems to be correlated with the size and shape of grains (along with intergranular area profiles).

## 4. Conclusion

In summary, in this article we have reported on the results obtained from the study of the elastic behavior of polycrystalline samples of superconducting $YBa_2Cu_3O_{7-d}$, prepared by the *sol-gel* method. The physical properties of our samples were determined using transport and magnetic measurements while their structural quality was checked by X-ray diffraction and scanning electronic microscopy. The elastic experiments were performed employing the standard *pulse-echo* technique by measuring the phase velocity and the attenuation of ultrasonic waves as a function of temperature. We have focused this study on the low temperatures interval (T < 200K) where there is no much information in the literature especially related to samples prepared following the *sol-gel* route.

Our results clearly indicate that the beginning of the superconducting transition occurs *before* the martensitic transformation associated with grain boundary properties and corresponding relaxation mechanisms. This transformation seems to appear as a consequence of changing temperatures acting on grains with different sizes and shapes. The structural response of those grains to the changing temperature generates a broad spectrum of stresses and strains between them (mainly around grain boundaries). This effect is more pronounced (hence observable) in non-homogeneous samples where the grain distribution in sizes and shapes is more broad. Our results clearly confirm the hysteresis in the attenuation at MHz frequencies, thus providing further evidence for the existence of the martensitic phase transition in these materials previously observed by other authors[9,10] in the kHz frequency range.

**Acknowledgements**

We gratefully acknowledge Brazilian agencies FAPESP and CNPq, and Uruguayan agency PEDECIBA for partial support.

**References**

[1] R. Truell, C. Elbaum, and B. Chick, in *Ultrasonic methods in Solid State Physics*, Academic Press, Inc.; 303, New York (1969).

[2] W. C. Wu and J. P. Carbotte; Phys. Rev. B60, 14943 (1999).

[3] A. K. Yahya and R. Abd-Shukor, Superconductor Sci & Technology 15, 302 (2002).

[4] F. Y. Li *et al*., Phys. Rev. B65, 132517 (2002).

[5] C. Lupien *et al*.; Phys. Rev. Lett. 86, 5986 (2001).

[6] J. Moreno and P. Coleman; Phys. Rev. B53, 2996 (1996).

[7] R. S. Kumar, H. Kohlmann, B. E. Light, A. L. Cornelius, V. Raghavan, T. W. Darling, and J. L. Sarrao; Phys. Rev. B69, 14515 (2004).

[8] Robert Joynt and Louis Taillefer; Rev. Mod. Phys. 74, 235 (2002).

[9] X. Cheng *et al*. J. Phys. C 21, 4603 (1988)

[10] C. Chen et al., Phys. Status Solidi A 113, K85 (1989)

[11] F. Fujita in *Physics of new materials*; 2nd.edition; Springer Series in Materials Science; v.27; Springer – Verlag, Berlin (1998).

[12] R. Pérez-Sáez *et al.,* Phys. Rev. B57, 10, 5684(1998).

[13] J. Szente *et al*., Phys. Rev. B37, 14, 8447 (1988).

[14] M. Fukuhara, Phys. Rev. B <u>69</u>, 224210, 1 (2002).

[15] G. Cannelli *et al*., Phys. Rev. <u>B45</u>, 931 (1992).

[16] M. P. Pechini, *Method of preparing lead and alkaline earth titanate and niobates and coating methods using the same to form a capacitor*, U. S. Patent, no 3.330.697 (1967).

[17] B. I. Lee and E. J. A. Pope, *Chemical Processing of Ceramics*; Ed. Marcel Dekker Inc.; New York (1994)

[18] H. M. Rietveld, J. Appl. Cryst. <u>2</u>; 65 (1969).

[19] F. Rodriguez-Carvajal, FULLPROF 98, Version 3.5, Dec. 1997, Laboratoire Leon Brillouin, CEA-CNRS (1997).

[20] R. Truell, C. Elbaum, and B. Chick, in *Ultrasonic methods in Solid State Physics*, Academic Press, Inc.; 53, New York (1969).

[21] E. Papadakis, *The measurement of Ultrasonic Velocity*; in: *Physical Acoustics*, Vol. XIX, Academic Press Inc. New York (1990).

[22] A. Moreno; *Ph.D. Thesis: Estudo experimental e teórico do processo de difusão e de ressonância de "kinks" em linhas de discordâncias de metais fcc submetidas a tensões oscilatórias de baixa amplitude*; Universidade Federal de Sao Carlos, Brazil (1997).

[23] T. Laegreid *et al*., Physica C <u>153-155</u>, 1096 (1988).

[24] S. Bhattacharya *et al.*, Phys. Rev. <u>B37</u>, 5901 (1988).

[25] J. Toulouse *et al*., Phys. Rev. B <u>38</u>, 7077 (1988).

[26] M. Xu *et al*., Phys. Rev. B <u>37</u>, 7,3675 (1988).

[27] L. Mamsurova *et al*., Physica C<u>167</u>, 11 (1990).

13# Figure Captions

**Figure 1** – Scanning electron microscopy (SEM) photography for sample M1 (left), and sample M2 (right).

**Figure 2** – Electrical resistivity as a function of temperature for two samples: (a) sample YBCO/M1; and (b) sample YBCO/M2. The inset shows a close-up of the double transition with a broadness typical of a granular system.

**Figure 3** – (a) Complex AC magnetic susceptibility as a function of temperature of sample YBCO/M1, for an external magnetic field H = 100 Oe, excitation AC magnetic field h = 1.0 Oe, and AC frequency $f$ = 1.0 kHz; (b) magnetization as a function of temperature of sample YBCO/M2 for an external magnetic field H=100 Oe.

**Figure 4** – (a) The temperature dependence of attenuation of longitudinal ultrasound waves with a frequency of 5MHz for a complete thermal cycle (cooling-down and warming-up processes) of sample M1; (b) velocity of longitudinal ultrasound waves with a frequency of 5MHz for a complete thermal cycle (cooling-down and warming-up processes) of sample M1. No hysteretic behavior is observed.

**Figure 5** – The temperature dependence of velocity and attenuation of longitudinal ultrasound waves with a frequency of 5MHz for a complete thermal cycle of sample M2.

**Figure 6** – Attenuation of longitudinal ultrasound waves with a frequency of 5MHz for the low-temperature interval (cooling-down and warming-up processes) of sample M2.

**Figure 7** - Comparison between the behaviors of velocity, magnetization (a.u.) and resistance (a.u.) as a function of temperature for the low-temperatures interval, for sample YBCO/M2 (hysteresis cycle).

**Figure 8** –Obtained results for M2 sample, for the velocity and attenuation at 5 MHz, superimposed with magnetic and transport results, for the warming-up process. The attenuation curve has been decomposed into four basic peaks. The meaning of each peak is explained in the text.

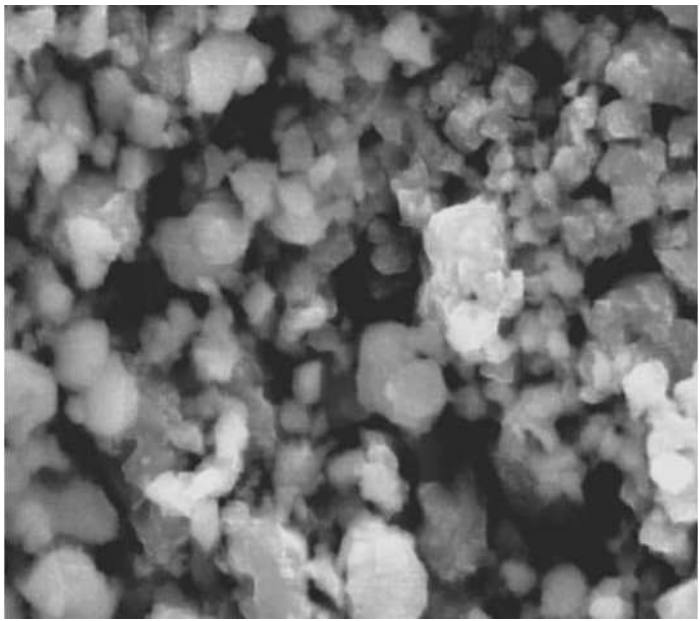 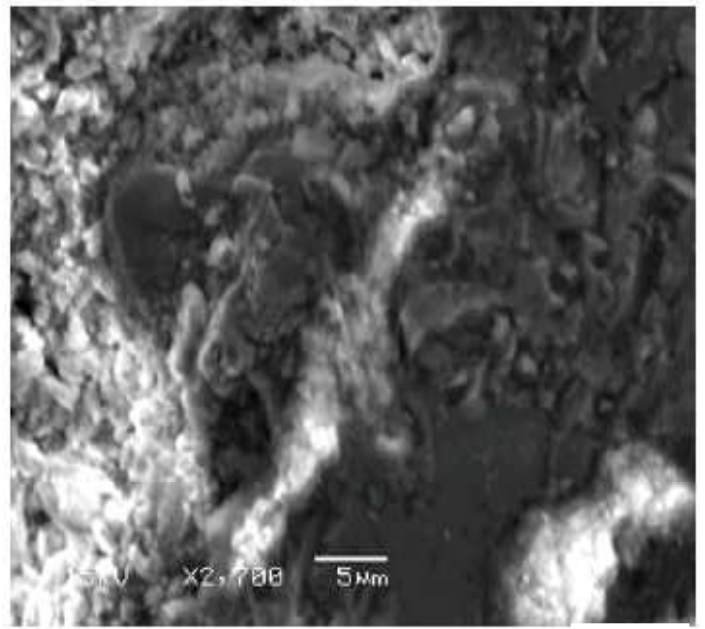

5μm  5μm

Figure 1; Stari *et al*.

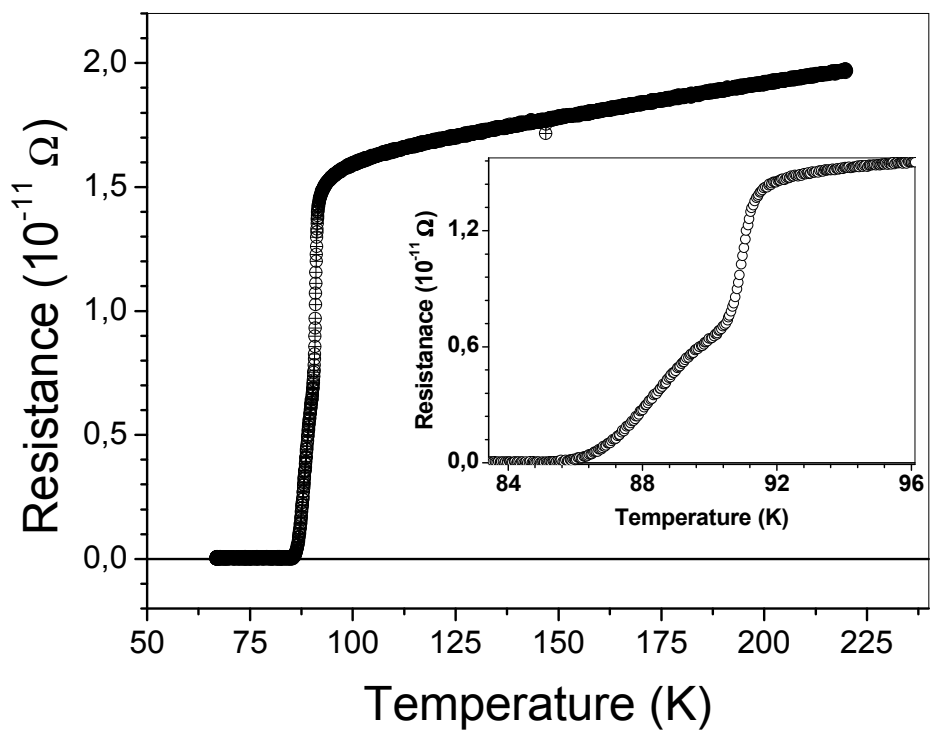

(a)

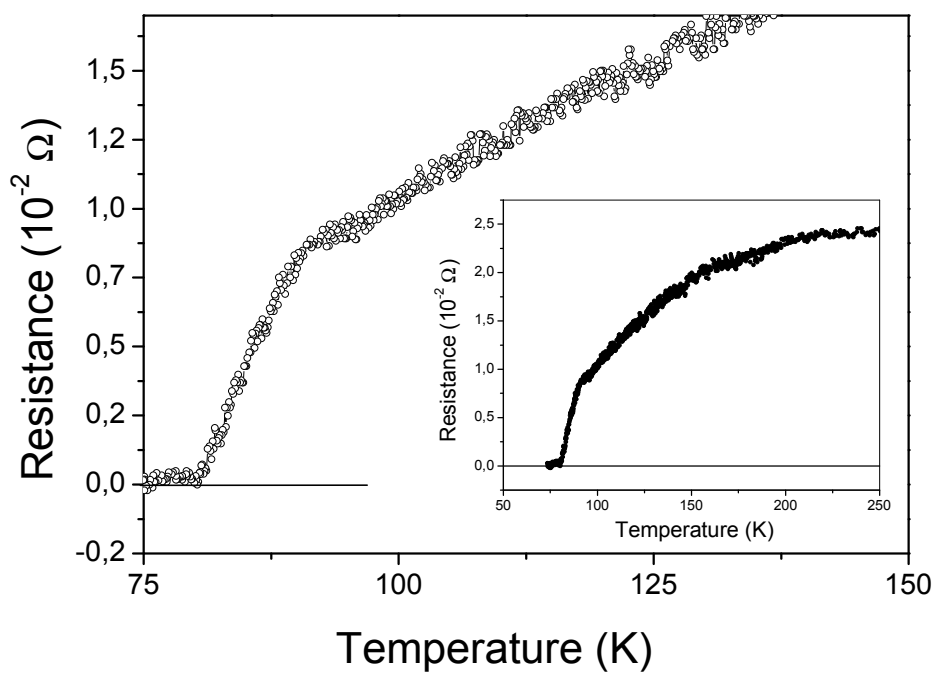

(b)

Figure 2; Stari *et al*.

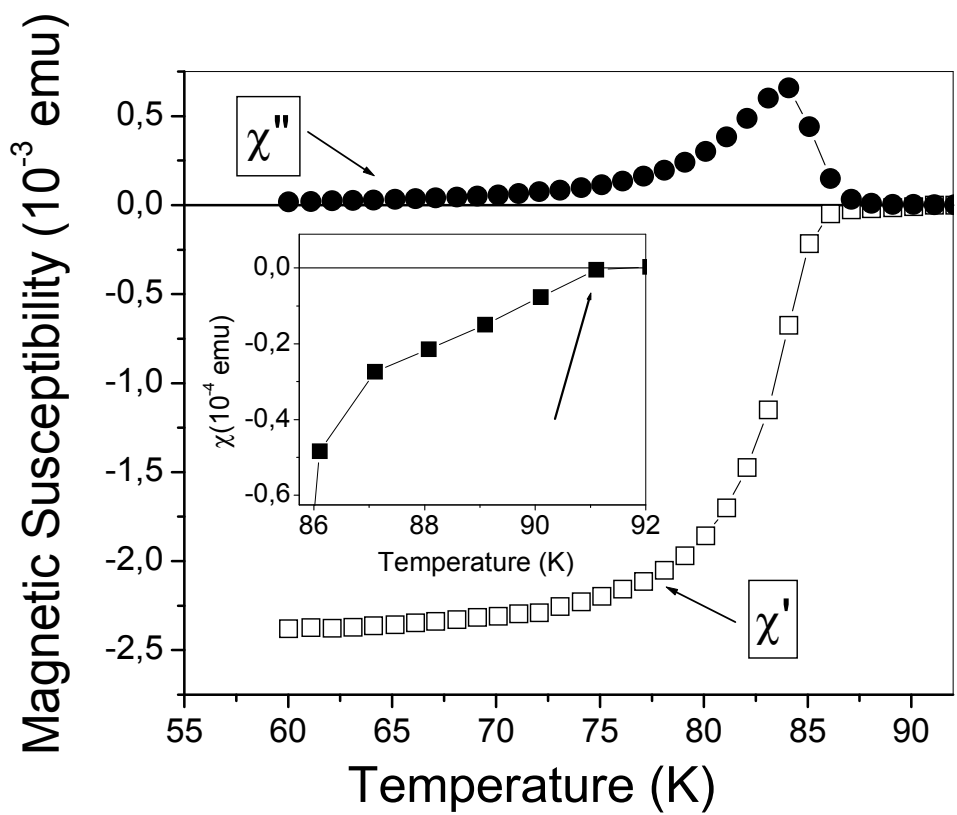

(a)

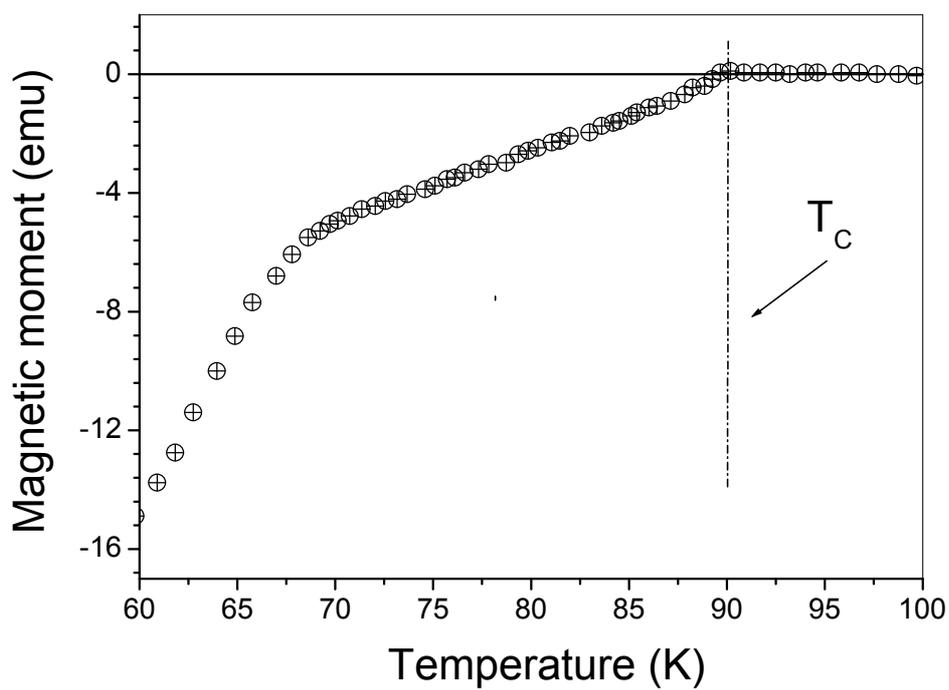

(b)

Figure 3; Stari *et al*.

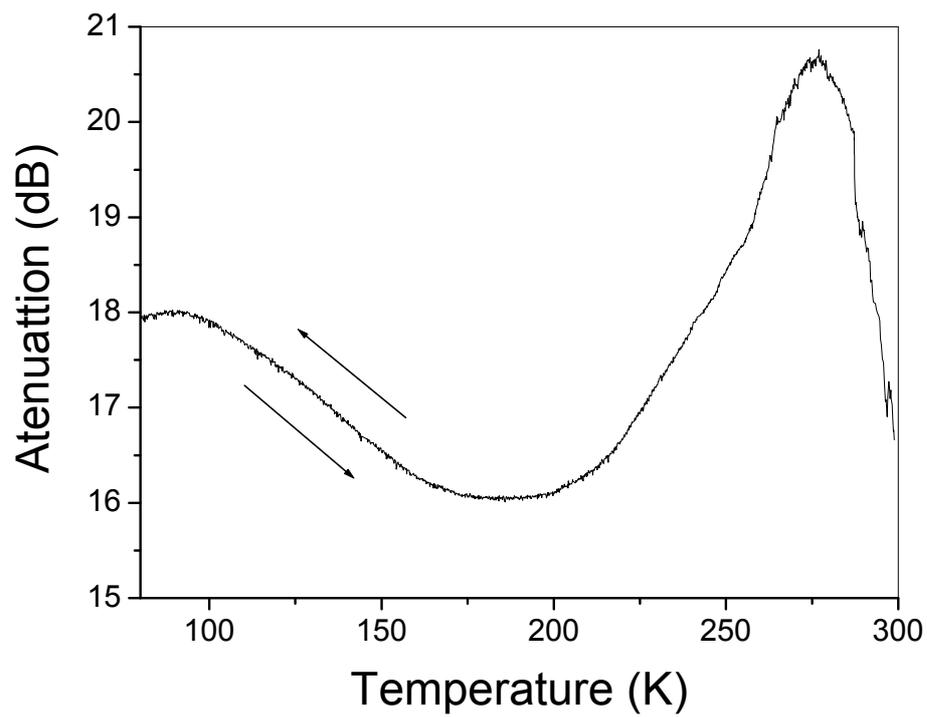

(a)

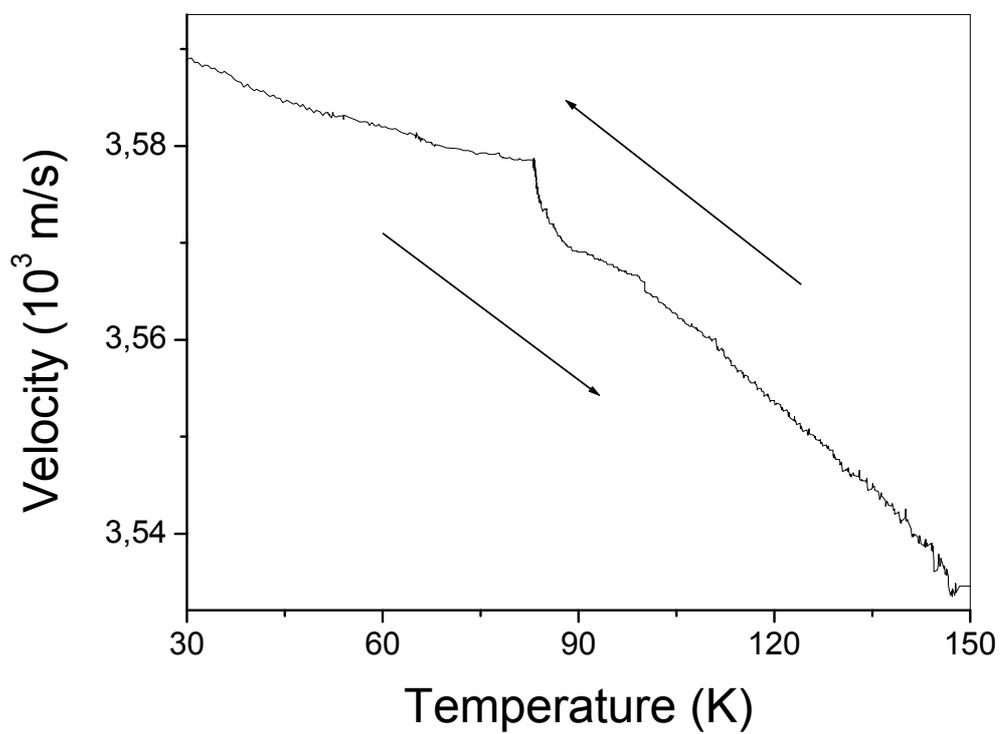

(b)

Figure 4; Stari *et al*.

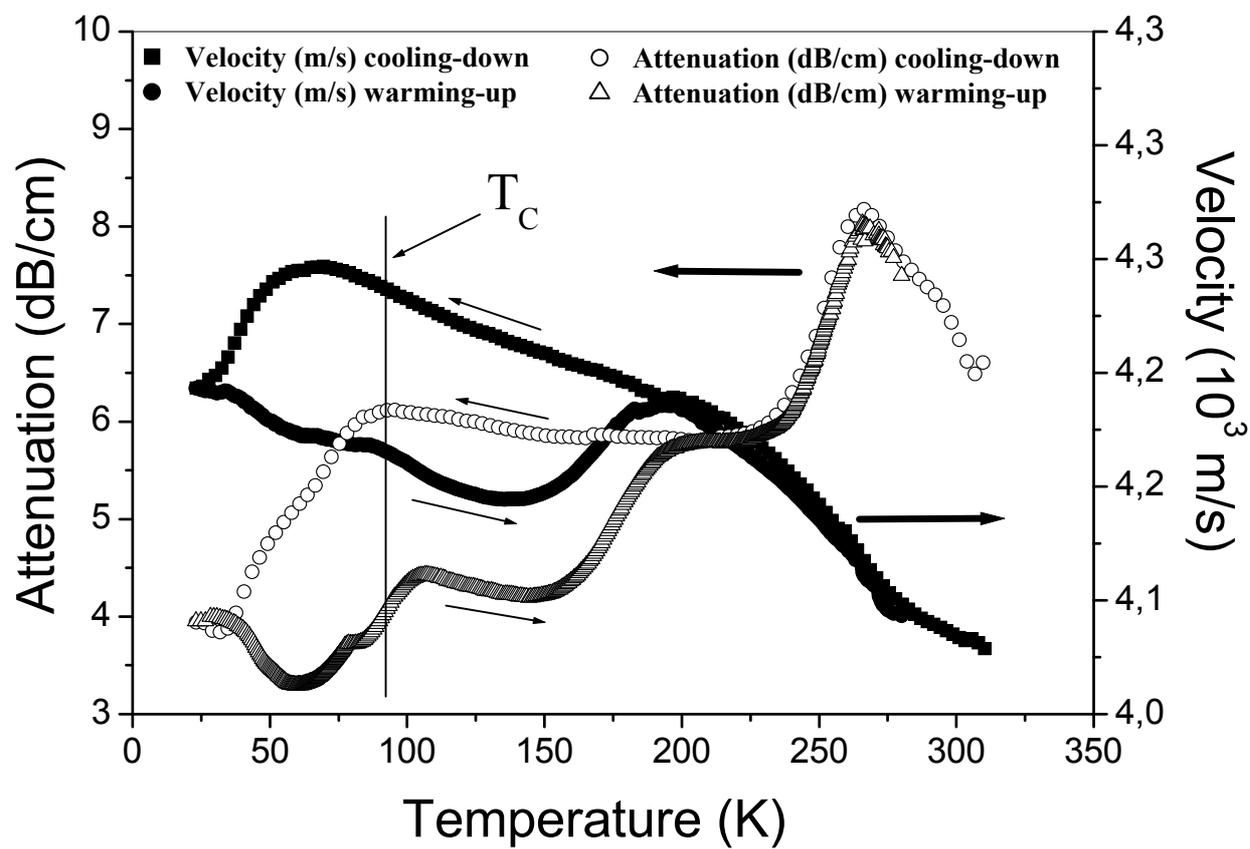

Figure 5; Stari *et al*.

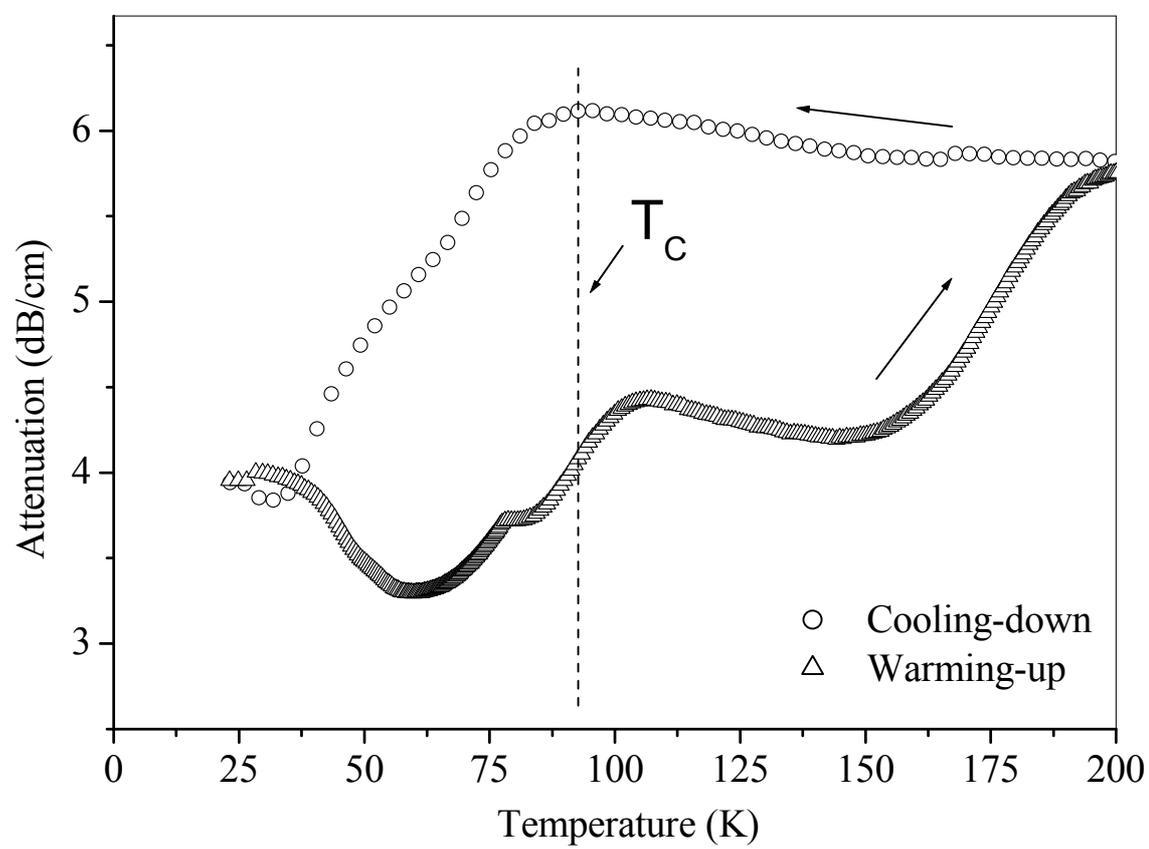

Figure 6; Stari *et al*.

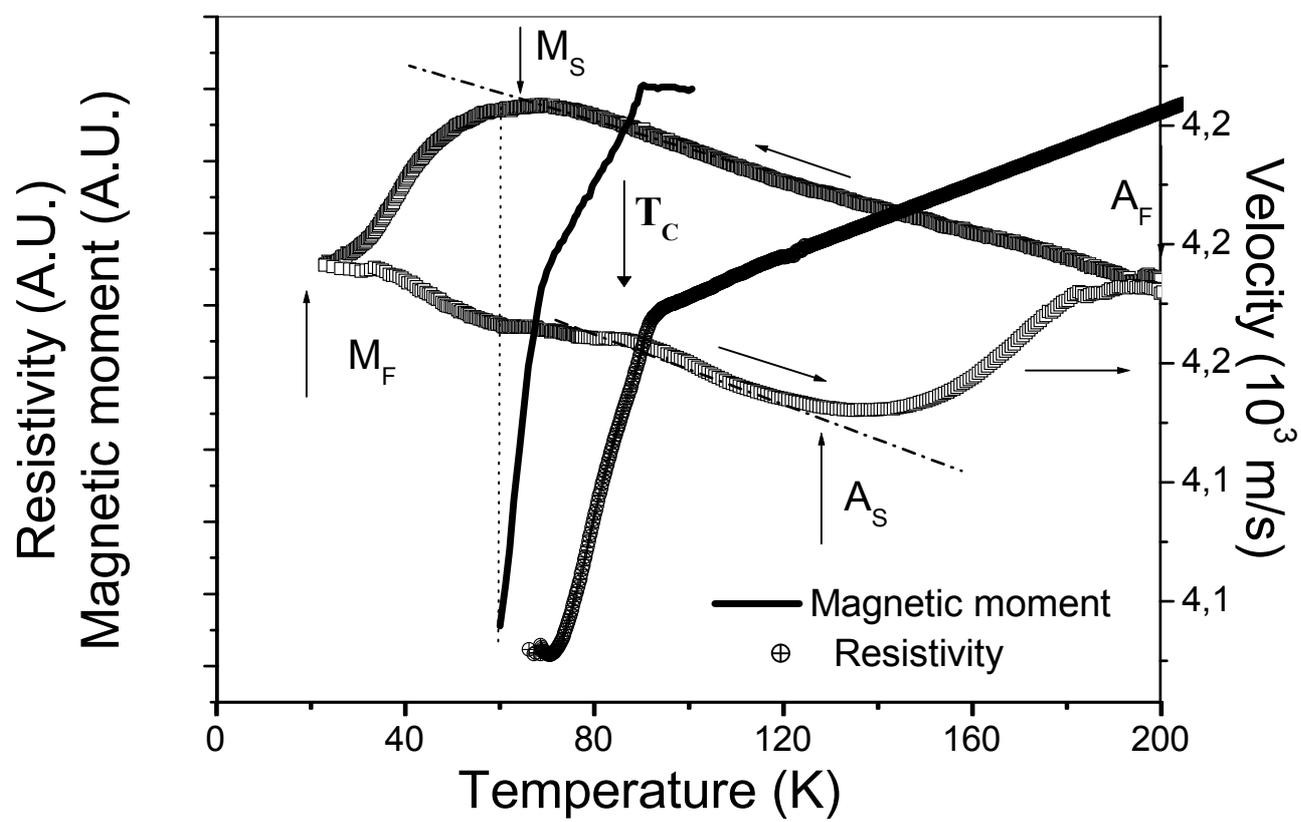

Figure 7; Stari *et al.*

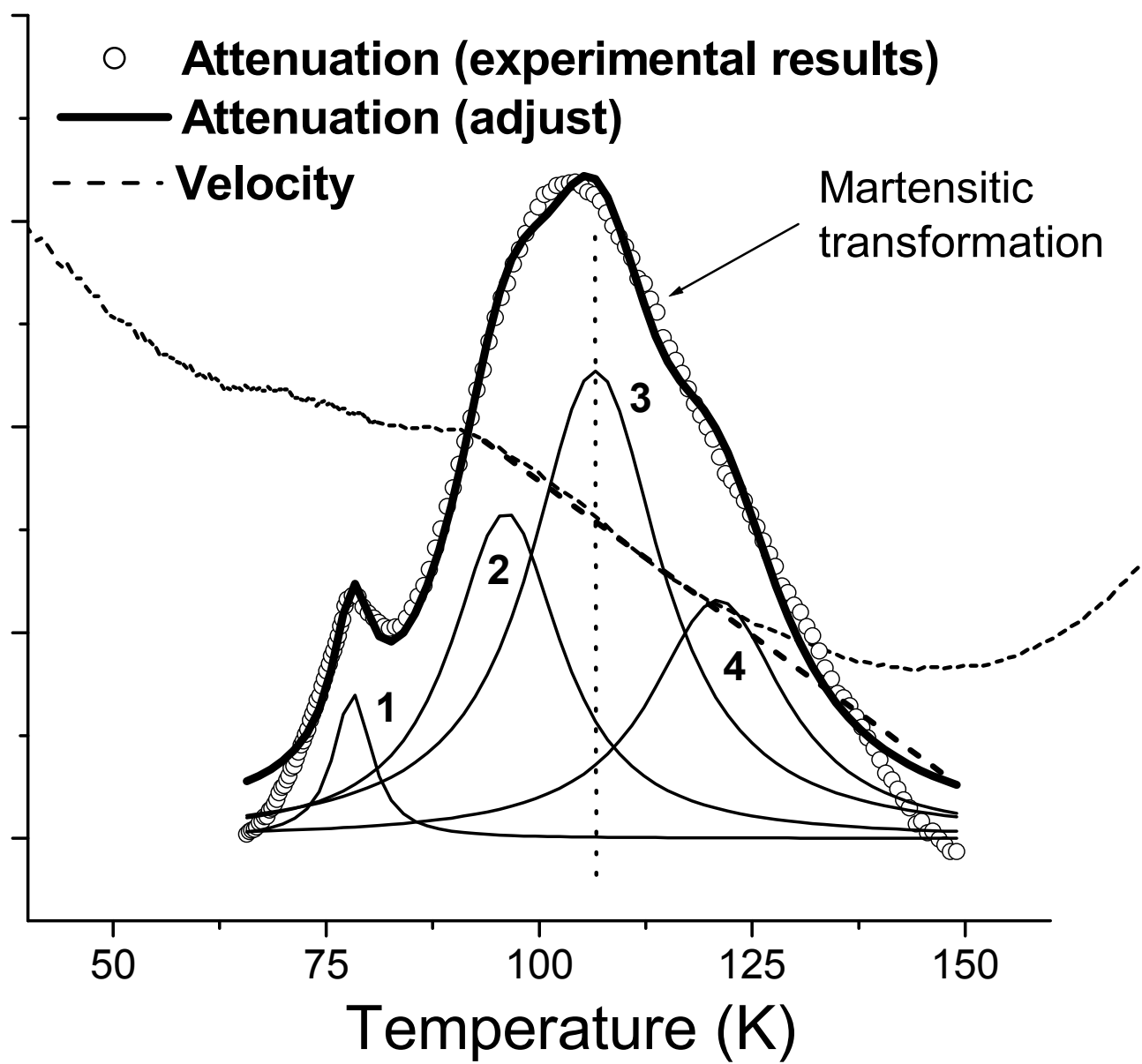

Figure 8; Stari *et al.*

**Table 1** - Fitting parameters of the Lorentz-like behavior for the four peaks of Fig. 8.

| Peak | A (K dB /cm) | $\omega$ (K) | $T_o$ (K) |
|---|---|---|---|
| 1 | 1.50 | 5.41 | 78.1 |
| 2 | 9.80 | 15.8 | 96.1 |
| 3 | 16.6 | 18.6 | 106.6 |
| 4 | 8.71 | 19.1 | 120.9 |

Table 1; Stari *et al*.